\documentclass[10pt,a4paper,twoside]{article}
\usepackage{epsfig}
\usepackage{baltlat6}
\usepackage{array}
\usepackage{epstopdf}
\begin{document}
\ \
\vspace{0.5mm}
\setcounter{page}{1}
\titlehead{Baltic Astronomy, vol.\, 22, 1--13, 2013}
\titleb{On the nature of V2282~Sgr}

\begin{authorl}
\authorb{R. Nesci}{1}
\authorb{C. Rossi}{2}
\authorb{A. Frasca}{3}
\authorb{E. Marilli}{3}
\authorb{P. Persi}{1}
\end{authorl}
\begin{addressl}

   \addressb{1}{INAF-IAPS, via Fosso del Cavaliere 100, 00133 Roma, Italy; roberto.nesci@iaps.inaf.it}
         \addressb{2} {Dep. of Physics, Univ. La Sapienza, Roma, Italy; P.le A. Moro 2, 00185 Roma, Italy; corinne.rossi@uniroma1.it}
             \addressb{3}{INAF-OACT, Via S.Sofia 78, 95123 Catania, Italy; antonio.frasca@oact.inaf.it}
\end{addressl}

   \submitb{Received: 2013 September 15; accepted .... }

% \abstract{}{}{}{}{} 
% 5 {} token are mandatory
 
\begin{summary}
The star V2282~Sgr is positionally consistent with a strong {\it Chandra} X-ray and a {\it Spitzer}/IRAC MIR source. We derived its long term $I$-band light curve from the photographic archives of the Asiago and Catania Observatories, covering the years from 1965 to 1984. CCD $R_C$ photometry in Summer 2009 was re-analyzed. Optical spectra were secured at Loiano Observatory in 2011 and 2012. 
J H K photometry, obtained from several experiments in different epochs was compared and the Spitzer images were re-analyzed.

V2282~Sgr was found to be irregular variable in all wavelengths.  Spectroscopically, it shows strong emission features (H Balmer lines, [{N}\small{II}]\,6584\AA~ and [{O}\small{III}]\,5007/4959\AA) while the {Na}\small{I} D doublet is very strong, indicating a circumstellar envelope. 
A single thermal energy distribution cannot reproduce the observed SED, while it can be explained as the sum of a G-type star plus a variable circumstellar disc, which mimics a class 0/I object. Most likely, V2282~Sgr is a 1-2 $M_{sun}$ mass pre main sequence star with an accretion disk.
\end{summary}

   \begin{keywords} stars: pre-main-sequence -- stars: protostars -- stars: variables: T Tauri--
                stars: individual, V2282 Sgr \end{keywords}
%
%________________________________________________________________
%% \resthead is the RUNNING TITLE at top of the pages
\resthead{On the nature of V2282~Sgr}{R. Nesci et al.}

\sectionb{1}{INTRODUCTION}
The M20 (Trifid) nebula is a spectacular star forming region of the Milky Way, projected in the sky over the open star cluster NGC~6514. 
Its distance has been estimated by several authors, ranging from 1.4 kpc (Ogura and Ishida 1975 and references therein) 
up to 2.7 kpc (Cambresy et al. 2001).  Maybe due to its larger distance, this nebula has been less studied than the well known Orion or Chamaeleon regions. 
Actually, after the systematic works of Maffei with photographic plates (Maffei 1959, Maffei 1963), no extensive optical search for variable stars was 
performed: the General Catalog of Variable Stars (Samus et al. 2012) lists only 4 nebular variables within 7 arcmin from the nebula center, against the 
several hundreds of variables known in Orion and Chamaeleon.

In the last decade a series of paper has been devoted to the study of M20, taking advantage also of space born instruments in the X-ray and far infrared wavebands
(Rho et al. 2004; Rho et al. 2006; Yusef-Zadeh et al 2005;  Cambresy et al. 2011).

The variable star V2282~Sgr, in the M20 nebula, was discovered by Lampland (1919, 1923) with a variability range of about one magnitude in the photographic 
blue and B$\sim$16\,mag at maximum. Three slitless grating spectra of the nebula were obtained by Herbig (1953) showing that the variable was of G-K spectral type, but the 
strong emission from the nebula masked the possible emission lines from the star. The star variability was studied for a few years by Maffei (1959, 1963) 
who classified it as an irregular variable and reported a variation amplitude of about 0.5 mag in the photographic infrared band.

The star was later listed as variable in Kukarkin et al. (1968) but its coordinates  were not enough accurate (RA=17:56:13 DEC=$-23$:02.9, 1900, corresponding to 
18:02:17.2 $-23$:03:00, J2000). Probably due to its inaccurate position, the star was completely forgotten in the following years. 
The actual position of the star, derived by us from a comparison of the finding chart published by Maffei (1959) and Herbig (1953) with the POSS plate, is 
instead RA=18:02:16.8 DEC=$-23$:03:47 (J2000). We informed about this the SIMBAD database which is now updated.

Due to its location in the sky inside the M20 nebula, the star is not present in the DSS-based USNO-B1 catalog. It is however present in the GSC2.3.2 catalog, 
albeit only in the $N$ band; it is well detected in the 2MASS (Cutri et al. 2003)  and DENIS (DENIS Consortium, 2005) surveys.

Rho et al. (2004) found the star positionally consistent with the soft X-ray source CXOM20 180216.8-230347 detected by {\it Chandra}. This source showed no 
variability in X-rays during their 16-hr observing run, with a flux level of 0.167 c/s. In the mid-infrared survey of the M20 nebula made by Rho et al. (2006) with the {\it Spitzer} telescope there is a source, positionally consistent with V2282~Sgr, classified 
by them as a proto-stellar source (Class 0/I n.16) on the basis of a MIR color-color diagram, but no comment was made in that paper about its possible optical counterpart. 

A nebular variable (either of the T Tau, RW Aur or T Ori type) is believed to be a pre main sequence (PMS) star already partially stripped of its originating circumstellar envelope (see, e.g. Glasby, 1974). This seems at odds with the classification as a Class 0/I object, which is a still deeply embedded source.
For this reason we decided to explore its nature in more detail, looking at old photographic plates and putting together the recent data  from ground- and space-based telescopes.

%                                     Two column figure (place early!)
%______________________________________________ Fig.1
   \begin{figure*}
%   \centering
 \includegraphics[width=13cm]{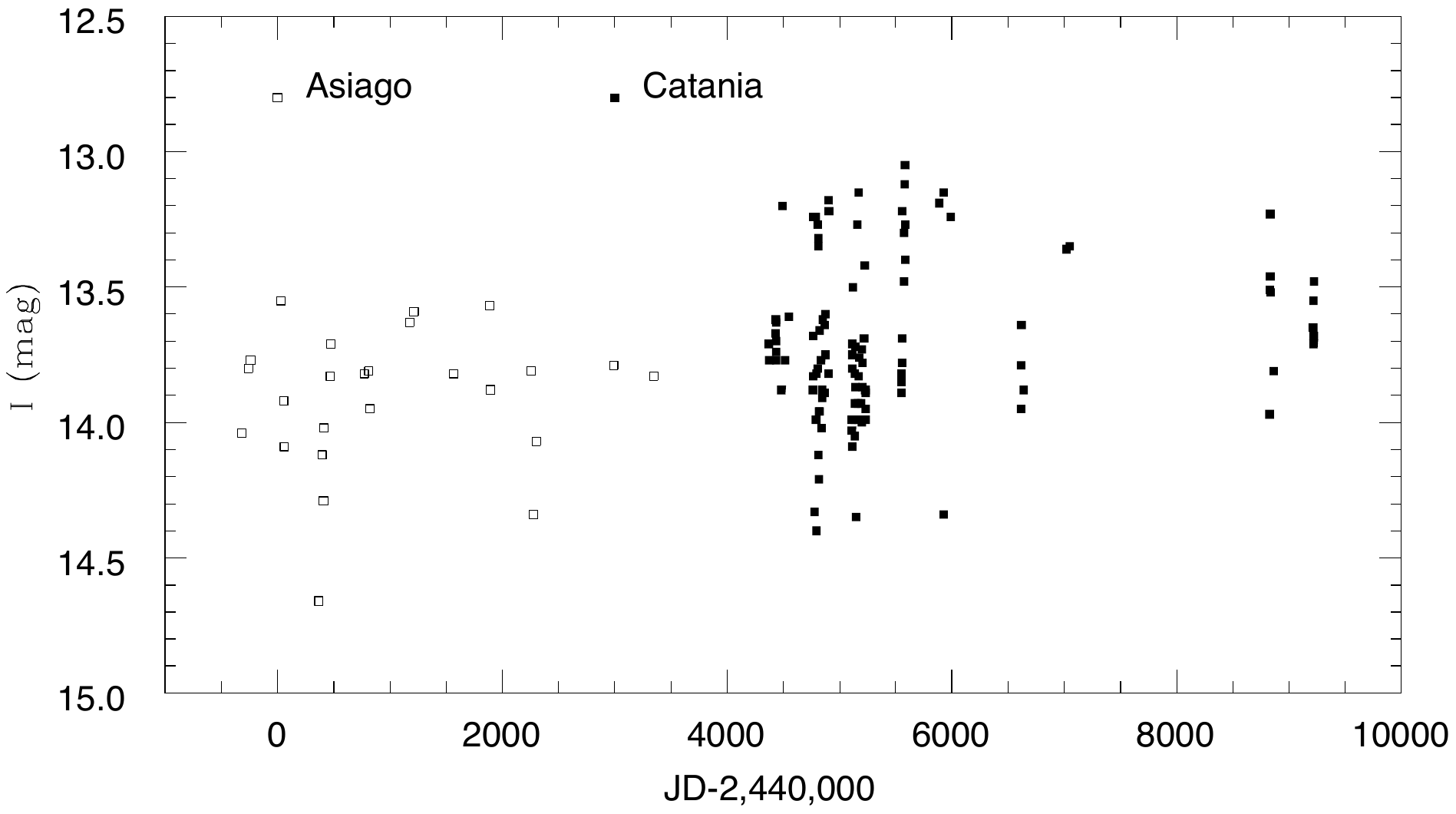}
     \captionb{1}{Light curve in the $I$ photographic band of V2282~Sgr from the Asiago (open squares) and Catania (filled squares) archives. 
Due to its southern position in the sky, the coverage shows a marked seasonal sampling.}
    \end{figure*}

\sectionb{2}{DATA SETS}

\smallskip

2.1 Photographic photometry

\smallskip

We measured photographic far-red magnitudes of V2282~Sgr on 25 plates of  the Asiago Observatory 65/92/215-cm Schmidt telescope, taken after the publication of the work by Maffei (1963). The plates were photographed with the sensitized Kodak I-N emulsion and the RG650 filter, a combination that matches very well the Cousins $I_{\rm C}$ photometric band. These plates cover a time span of ten 
years, from 1967 07 09 to 1977 07 24, and they have not been used previously for investigation of this variable.
The plates were digitized with an Epson 1680Pro scanner in the transparency mode, at 1600 dpi, corresponding to a scale of 1.5 arcsec/pixel 
(see Nesci et al. 2006 for more details on the scanning and photometric procedures). 

Aperture photometry was performed with the task \textsc{apphot} of the IRAF \footnote{IRAF is distributed by the National Optical Astronomy Observatory, 
which is operated by the Association of Universities for Research in Astronomy, under cooperative agreement with National Science Foundation.} software
using a 2 pixel radius. A reference sequence of 21 nearby stars 
was selected from the GSC2.3.2 catalog in the $N$ band, ranging from $N=12.0$ to $N=14.5$.

For each plate,  a linear fit of magnitudes of the comparison stars was adopted to interpolate the magnitude of the variable. We checked that the results were not significantly dependent on the assumed aperture radius and sky background.

In the archive of the Catania Astrophysical Observatory we found 109 useful plates (I-N emulsion and RG650 filter) taken by S. Cristaldi with the 41/61/121 cm Schmidt 
telescope at Serra La Nave (Mt. Etna, 1735 m above sea level) between 1980 05 12 and 1994 07 05. The large majority (90) of the plates were secured between 1980 and 1983, with only a few plates scattered in the later years. The plates were digitized at the Catania Observatory, and photometry was made using the same comparison stars as at Asiago. Unfortunately there was no temporal overlap with the Asiago plates. 

The photometric accuracy of our magnitudes ranges from 0.08 mag at $I=13$ to 0.18 mag at $I=14.5$: fainter stars have larger dispersion, as expected. The Catania plates, despite their smaller angular resolution, are not worse than the Asiago ones. The reason might be that the field was observed at Asiago at larger airmasses.  
The rms deviation of the values of $N$ magnitude for V2282~Sgr was 0.25 mag for the Asiago and 0.34 mag for the Catania datasets, substantially larger than for the faintest comparison star.

The resulting light curve based on the Asiago and Catania observations is shown in Fig.~1. The star shows substantial variations of brightness, up to 1 mag, with fast flares and dips, with an average level $N=14.0$ mag. 
A systematic difference in the upper envelope values of the Asiago and Catania observations seems to be present, but we feel it is just due to a denser time sampling of the Catania data. If real, an overall secular brightening of the source could be the most likely physical explanation.

\smallskip

2.2 CCD images

\smallskip

A large number of images of M20 were taken at the Cornero Observatory\footnote{http://www.cdso.it} with a 30cm F/6.5 Schmidt-Cassegrain telescope. Most of images were obtained with a CCD camera with the$R_{\rm C}$ filter during a dedicated observational season from 2009 June to October, on 19 different nights. A preliminary version of the light curve is present in the web site maintained by R. Behrend of the Geneva Observatory (Cornero 2009).
We made aperture photometry of these images using 12 reference stars with red (F-band) magnitudes from the GSC2.3.2 catalog. The scatter of the individual reference stars from the linear fit between instrumental and nominal magnitudes was rather large (0.24 mag) and systematic for each star, so we made an internal re-calibration of the sequence using a night with 100 frames. 
The scatter of the instrumental magnitudes of the comparison stars from this re-calibrated sequence was between 0.01 and 0.03 mag in all the other nights, supporting a very good internal accuracy of our photometry. No reliable intra-night variability was detected in the nights with long coverage. The resulting light curve is shown in Fig.~2. 

%----------------------------------------------------------- fig.2
   \begin{figure}
 %  \centering
  \includegraphics[width=8cm]{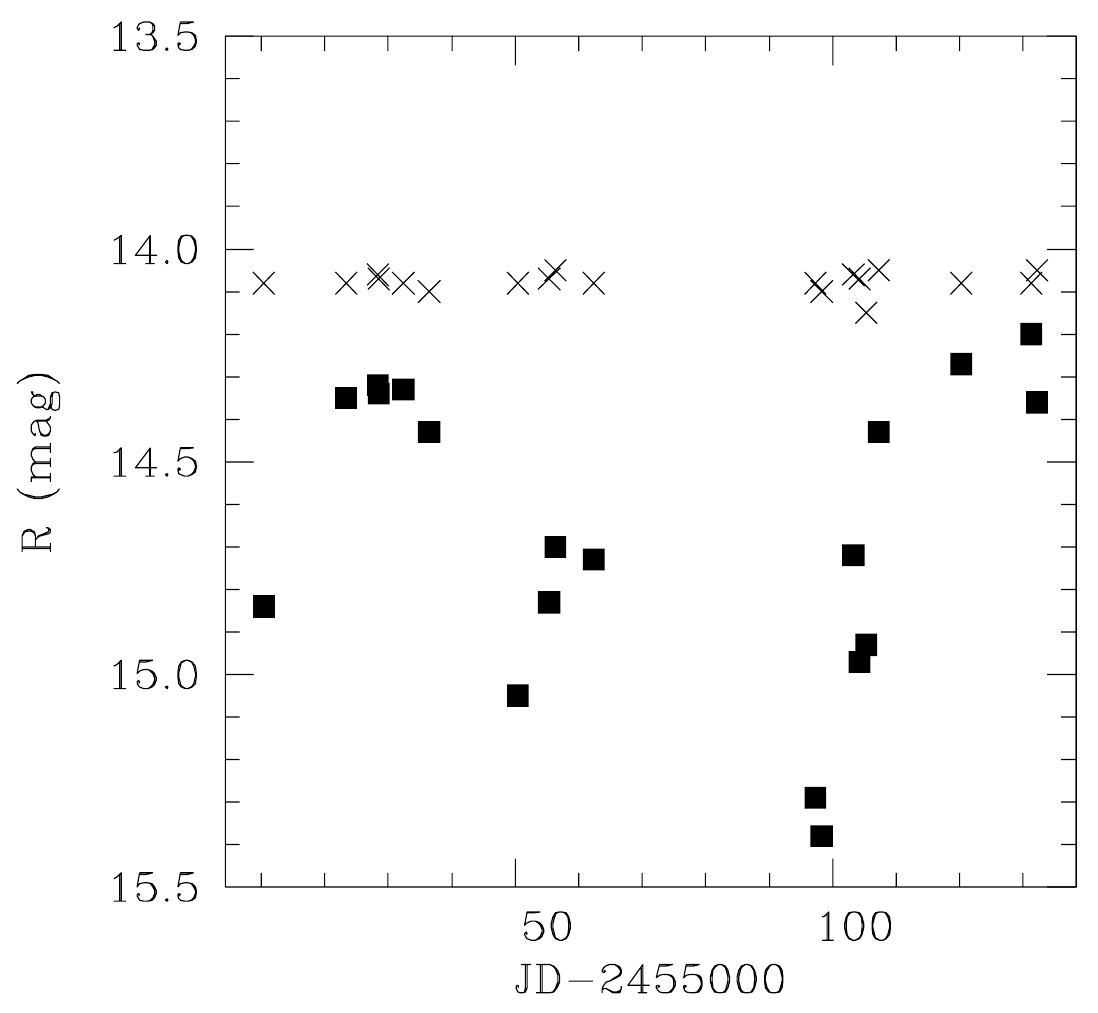}
      \captionb{2}{The light curve of V2282~Sgr in the summer 2009 from the CCD images. Filled squares are for the variable, crosses are for a comparison star of $F=14.07$. }
   \end{figure}
%
%______________________________________________________________

The star shows an oscillating behavior, again with an amplitude of about one magnitude. The distance between the local minima is about 50 days, but the time 
base is too short to allow any conclusion on a possible periodicity on this time scale. The time interval between a local minimum and maximum may be as short as 10 days, 
so that the loose time sampling of historic plates may underestimate the actual variability range.

Finally,  with the BFOSC instrument at the Loiano 1.52 m telescope we obtained the single $B$, $V$ and $R$ magnitudes in 2011 and one $R$ magnitude in 2012. To derive the magnitudes we used a subsample of the 
previously defined comparison stars (due to a smaller field of view).  The rms dispersions around the best fit calibration lines were 0.03 mag. The star was at  $B=17.60$, $V=16.31$, $R=14.70$ in 2011 and $R=14.40$ in 2012; both $R$ values are within the variability range of the light curve 
of 2009.

The magnitude dips of about 0.5--1.0 mag that we observed both in the $R$ (Figure~ 2) and in the photographic $I$ (Figure~1) light curves 
are typical of nebular variables and are reminiscent of the dimming episodes lasting a few days found by Frasca et al. (2009) in some variable stars around the Orion Nebula Cluster, such as {AA~Ori}, {AB~Ori}, and  {V411~Ori} (see their Fig.~22).

\smallskip

2.3 Optical spectroscopy

\smallskip

We have obtained two spectra of V2282~Sgr (on 2011 08 02 and 2012 07 18) with the Loiano 152cm telescope and the BFOSC instrument: the exposure time was 1800s, using a 2" slit 
and grism 4, giving a sampling of 4 \AA/pixel (resolution about 12 \AA).
The spectra were reduced with the IRAF reduction package, following the standard procedures. Due to the large airmass no reliable flux calibration could be obtained.

Since the star is embedded in the surrounding M20 nebular emission, we first studied the intensity of the nebular emission lines visible in the two dimensional 
spectrum along the slit position (13'). In the extraction procedure of one-dimensional spectrum we tried several solutions for the aperture width and the position of the background to subtract.
As a check, we extracted with the same parameters the spectra of two other stars  present by chance in the slit, namely {2MASS J18020361-2303468} and {2MASS J18020707-2303490}.
In both cases the strong emission lines from the nebula and from the Loiano sky were completely removed by the extraction procedure. 

The one-dimensional spectrum of V2282~Sgr shows a number of emission lines  which are most probably formed in a region physically close to the star. 
The two spectra separated from each other by one year, look rather similar. The only remarkable difference is that the [{N}\small{II}]\,6584\AA~ line is not apparent in July of 2012. 
The average of our two spectra of V2282~Sgr is shown in Figure 3a and the H$\alpha$ spectral region is shown enlarged in Figure 3b. 
The emission lines in the spectrum are listed in Table 1,  together with their equivalent width ($EW$).

%%%%%%%%%%%%%%%%%%%%%  table 1 
\begin{table}
\begin{center}
\vbox{\footnotesize\tabcolsep=3pt
\parbox[c]{124mm}{\baselineskip=10pt
{\smallbf\ \ Table 1.}{\small\
Emission lines in the 2011 spectrum of V2282~Sgr. \lstrut} } % title of Table
\begin{tabular}{ c c c }        % centered columns (4 columns)
\hline\hline                 % inserts double horizontal lines
wavelength & ion & $EW$ (\AA) \hstrut\lstrut\\   % table heading 
\hline                        % inserts single horizontal line
 6584& [{N II}]& -5 \hstrut \\
 6563& H$\alpha$ & -20  \\
 6548& [{N II}]& -0.6:: \\
 5007& [{O III}] & -10 \\
 4959& [{O III}] & -5 \\
 4861& H$\beta$ & -16  \\
 4363& [{O III}] & -6:: \\
 4340& H$\gamma$ & -15 \lstrut \\
\hline                                   %inserts single line
\end{tabular}
 }
\end{center}
\vskip-8mm
\end{table}
%%%%%%%%%%%%

The strongest emission line is H$\alpha$, blended with the [{N}\small{II}]\,6584\AA~  line;  the $EW$ of the fainter nitrogen line of the same doublet, [{N}\small{II}]\,6548\AA~  is measured with low accuracy, as indicated by the two colons in Table 2.

%----------------------------------------------------------- fig.3
   \begin{figure}
%   \centering
  \includegraphics[width=13cm]{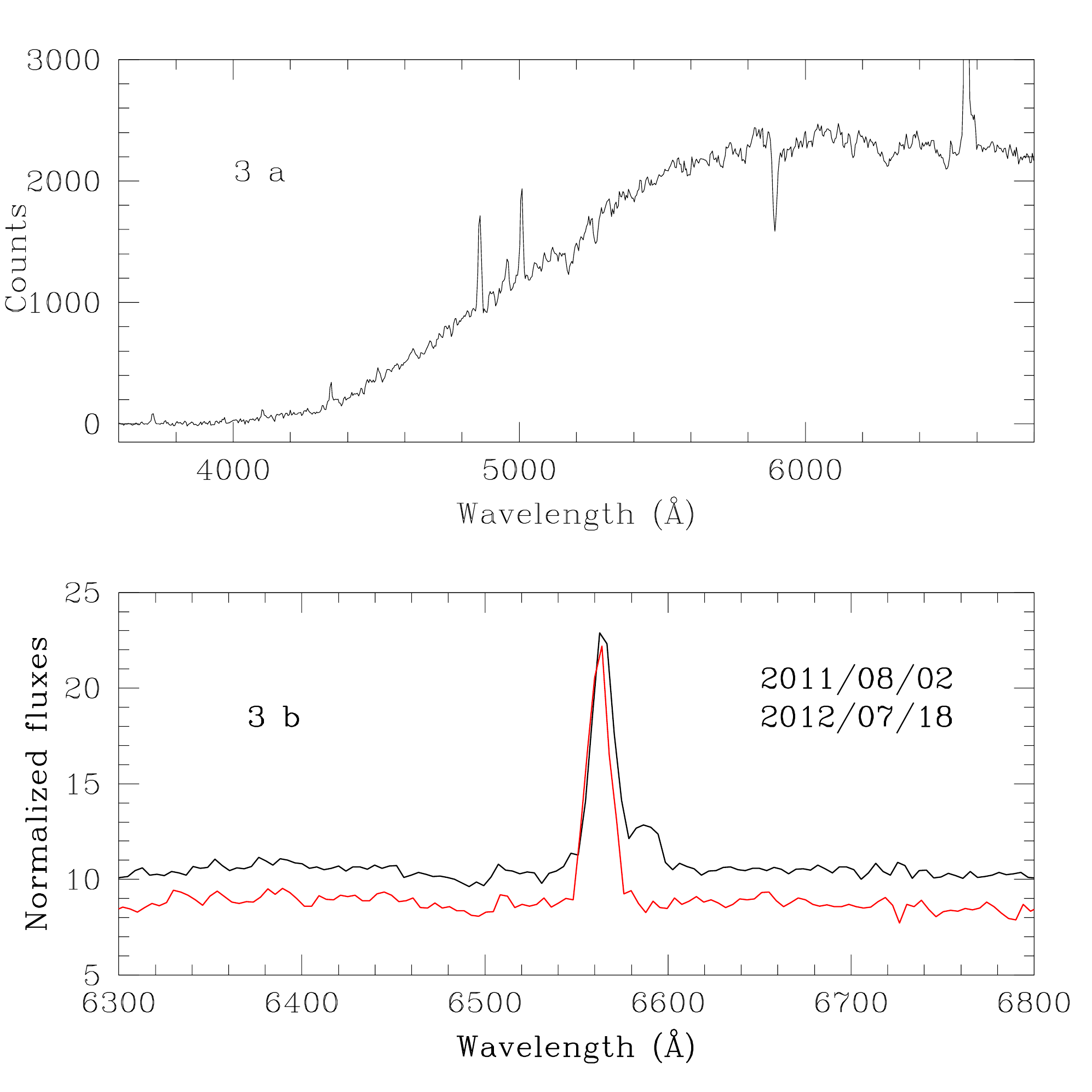}
      \captionb{3}{ Upper panel: the wavelength calibrated spectrum of V2282~Sgr, showing the emission lines of hydrogen, [{N}\small{II}]\,6584 \AA ~,  [{O}\small{III}]\,5007/4959 \AA, 
      and [{O}\small{II}]3727/29 \AA.~~
   Lower panel: enlargement of  the  spectra of V2282~Sgr taken in 2011 August (upper line) and 2012 July (lower line). The lack of the [{N}\small{II}] lines in 2012 is evident.}
      
%         \label{spectrum}
   \end{figure}

The  other Balmer lines are also present in emission,  as well as the [{O}\small{III}] doublet at 4959/5007 \AA~ and the [{O}\small{II}] doublet at 3727/29 \AA.  

None of the other emission lines often present in T Tau and RW Aur stars, like [{O}\small{I}]\,6300\AA,  {He}\small{I}\,5876\AA,  [{S}\small{II}]\,6717/30\AA~ doublet (Cohen \& Kuhi 1979), are detectable in V2282~Sgr, while they are present in the spectrum of the surrounding M20 nebula even around the star.
On the other hand, several emissions present in the  M20 nebula,  like the {He}\small{I}\,5876\AA~  and the  [{Ar}\small{III}]\,7136\AA~  lines disappear after the spectral extraction. 
Also all the Loiano night sky emission lines disappear from the extracted spectrum when the adjacent sky spectrum is subtracted. We are therefore 
confident of the reality of the emission lines detected in V2282~Sgr.

Among the absorption lines, the most prominent are the {Na}\small{I}~D doublet (6.5\,\AA), the {Mg}\small{I}b triplet (3.3\,\AA), the  blends of  several atomic lines centered at 5269 and 6497\,\AA, the {Ca}\small{I}\,4227\,\AA~ and the G band, indicating an intermediate G-K spectral type.
The {Na}\small{I}~D doublet is very strong, compared to the typical value (2.5\,\AA) of the main sequence or giants G-K stars, suggesting the presence of substantial interstellar/circumstellar matter, as expected in a PMS star.

From the absorption lines the spectral type is between G4 and K2: assuming K0, with ($B$--$V$)$_0$=1.0, we derive that the reddening  $E(B-V)$ is about 0.3, or larger for a spectrum of an earlier type.
This is in agreement with the estimate of the reddening towards M20 given by Kohoutek et al. (1999) and within the range of the literature values going from 0.23  to 0.4 (Ogura  \& Ishida 1975; Lynds et al 1985; Kohoutek et al. 1999).

\smallskip

2.4 Infrared data

\smallskip

In order to understand the nature of V2282 Sgr we have analyzed the near- and mid-IR images taken from different surveys and used published data. 
In the near-IR we found four observations taken at different epochs. In Table 2 we report the $J$, $H$ and $K$ magnitudes taken from 2MASS, DENIS and UKIDSS catalogs and from the observations reported by Rho et al. (2006) obtained with the Palomar 5 m telescope. 

%%%%%%%%%%%%%%%%%%%%%%   table2
\begin{table}[!b]
\begin{center}
\vbox{\footnotesize\tabcolsep=3pt
\parbox[c]{124mm}{\baselineskip=10pt
{\smallbf\ \ Table 2.}{\small\ Near Infrared Photometry.\lstrut}}
\begin{tabular}{c c c c c c c}        % centered columns (7 columns)
\hline\hline                 % inserts double horizontal lines
Date& $J$ & $H$ & $K$ & ($J-H$) & ($H-K_s$) & Note \hstrut\lstrut\\  % table heading 
\hline                        % inserts single horizontal line
 1998/06/14     & 12.74(0.03)  &  12.02(0.03)  & 11.54(0.03)  & 0.72   &  0.48  &  2MASS \hstrut\\
 1996/04/01     & 12.90 (0.08) &   ...         & 11.51(0.08)  &  ...   &  ...   &  DENIS \\
2004/08/23-26   & 12.99(0.02)  &  12.26(0.02)  & 11.93(0.02)  & 0.73   &  0.33  &  Palomar 200inch \\
2006/07/23      & 15.71(0.01)  &  14.21(0.01)  & 12.78(0.03)  & 1.50   &  1.43  &  UKIDSS \lstrut \\
\hline
\end{tabular}
}
\end {center}
\vskip-8mm
\end{table}

%%%%%%%%%%%%%%%%%%

The values in Table 2 allow to conclude that V2282~Sgr possessed no strong IR excess at the epochs of the 2MASS, DENIS and Palomar observations, while it was much fainter and redder at the UKIDSS epoch. In that season the source was much fainter also in the optical range, as 
shown in an image of the Cornero Observatory archive (2006 05 28);  by comparison with the Loiano images we derived for that date an upper limit $V\sim17.5$\,mag.

Figure 4 shows the color composite image obtained from the three UKIDSS frames in the $J$, $H$, and $K_s$ passbands: a very weak source ($K_s=14.25$\,mag), not visible in the 2MASS images, lies $\sim 3$ arcsec eastward of our star.

%----------------------------------------------------------- fig.4
   \begin{figure}
%\centerline{
\includegraphics[width=9.5cm]{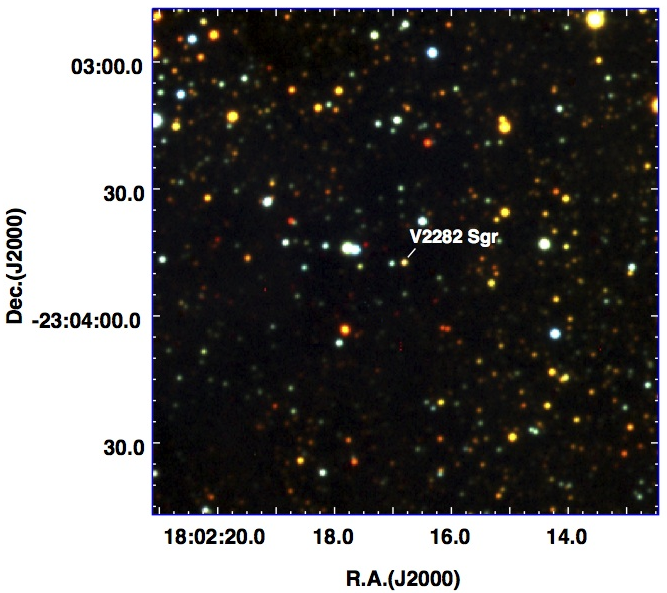}
      \captionb{4}{Color composite of V2282~Sgr from UKIDSS images in $J$, $H$, and $K_s$ passbands. The fainter star to the left from the variable, quoted in the text, is not resolved in the 2MASS image. }
%         \label{cartinaU}
   \end{figure}
In the ($J-H$)/($H-K_s$) plane, (Figure~5) we plotted positions of our star on three different observing dates. While in 1998 and 2004 the star is close to the locus of classical T Tau stars, in 2006 it shows a strong near-IR excess. This transient $K_s$-excess phenomenon has been observed in a few PMS stars 
inside the dark cloud {L~1003} in the {Cyg~OB7} association by Rice et al. (2012).
Near-IR variability is a common phenomenon in PMS stars: indeed Kaas (1999) used the $K_s$-band variability as a method to select young stellar objects in the Serpens cloud.

The {\it Spitzer}/IRAC images between 3.6 and 8.0 $\mu$m were taken in three different epochs as reported in Table 3: the observations of the program AOR 6049024 are
those reported by Rho et al. (2006). The star V2282 Sgr appears very bright in the four IRAC bands as shown in the color composite image of Figure~6 obtained from the 3.6 (blue), 4.5 (green) and 8.0 (red) $\mu$m images. The bright emission at 4.5 $\mu$m at south west of the star could indicate the presence of an H2 jet, because the 4.5 $\mu$m band is dominated by shocked H$_2$ emission lines (Noriega et al. 2004, Ybarra et al. 2009). We performed aperture photometry of the star with the \textsc{apphot} task of IRAF using the same aperture size of Rho et al. (2006). The results of our measurements  for AOR 6049024, reported in Table~3, are in agreement with those by Rho et al. (2006). The star is significantly variable, mainly at shorter wavelengths, as also reported by Robitaille et al. (2008), enforcing the evidence on the PMS nature of V2282~Sgr. The{\it Spitzer}/IRAC variability was also detected in 19  Class I/II T Tau stars in the Vela molecular cloud D (Giannini et al. 2009) and in the IC1396A dark globule (Morales et al. 2009).

The star V2282 Sgr is also present in the WISE catalog (Cutri et al. 2012) with a strong mid-IR excess (10.28 $\pm$0.05 mag in 3.4 $\mu$m; 9.07 $\pm$0.03 mag in  4.8 $\mu$m; 
5.50 $\pm$0.02 mag in 11 $\mu$m; 1.82 $\pm$ 0.04 mag in 22 $\mu$m) . In the 3-12 $\mu$m range, both the WISE and the {\it Spitzer} data are well fitted by a power law in the  $log(\lambda F_{\lambda})$ vs. $ log(\lambda)$ plane with very similar slopes; the WISE data are about 0.3 dex brighter than the {\it Spitzer} ones: this is true also for the point at 24 $\mu$m. Such difference is quite compatible with the variability range of the source. V2282 Sgr is present also in the MSX6C Infrared Point Source catalog, but with low reliability.
% calibration 8.18E-15; 2.415E-15; 6.515E-17; 5.090E-18 W/cm2/micron
% ovvero 306.68; 170.66; 29.044; 8.284 Jy.
% conversione da W/cm2/micron a erg/s/cm2/micron moltiplicare per E+07
% quindi 8.18E-08; 2.415E-08; 6.51E-10; 5.090E-11 erg/s/cm2/micron

%----------------------------------------------------------- fig.5
\begin{figure}
%\centering
\includegraphics[width=10cm]{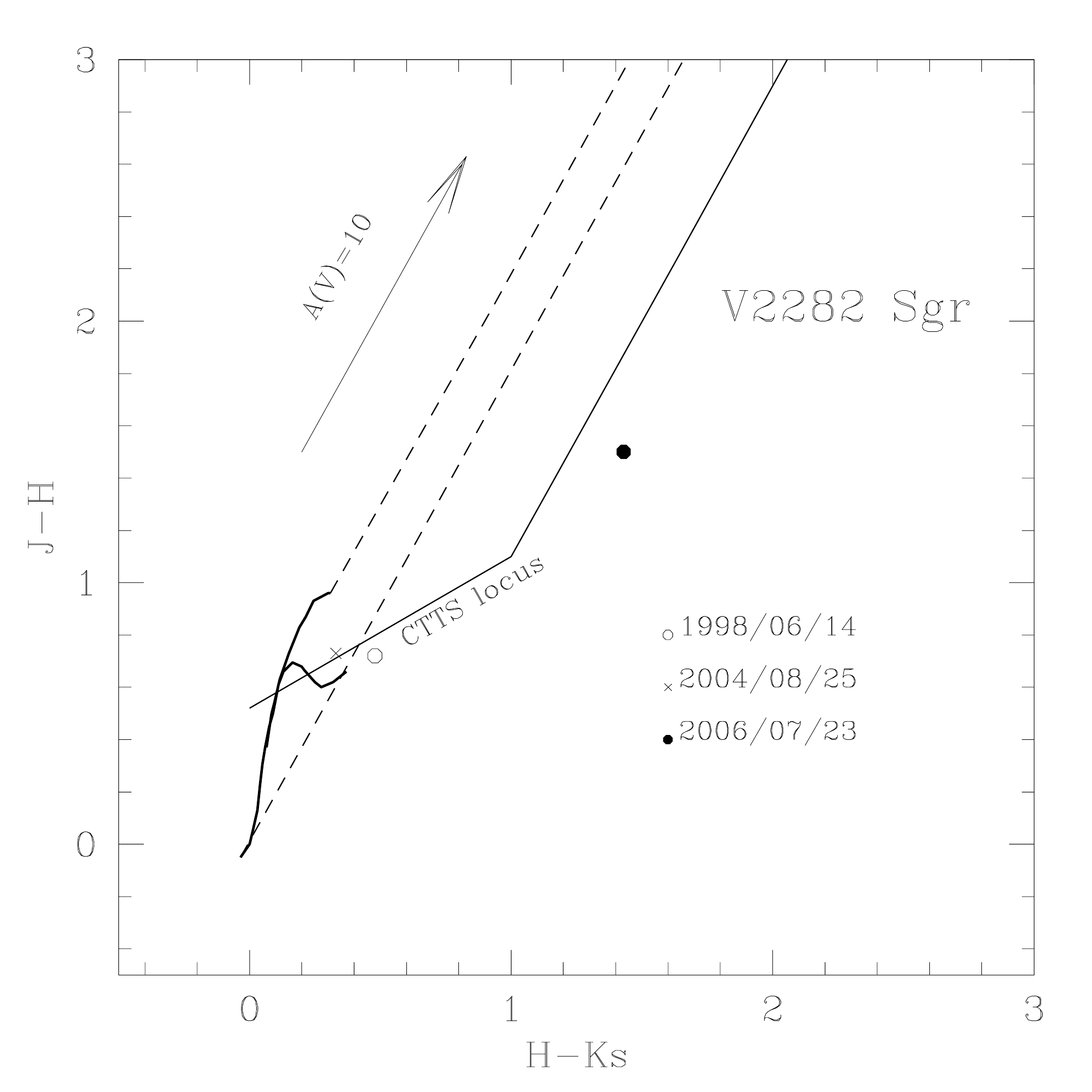}
\captionb{5}{The position  of V2282~Sgr in the ($J-H$),($H-K_s$) plot at three epochs from the literature data: the main sequence locus and the reddening line are 
shown. It is evident that the source was redder when it was fainter, as expected in case of matter ejection.}
%\label{ctts}
\end{figure}
%
%______________________________________________________________
%%%%%%%%%%%%%%%%%%%%%%%%%% table 3 

\begin{table}[!t]
\begin{center}
\vbox{\footnotesize\tabcolsep=3pt
\parbox[c]{124mm}{\baselineskip=10pt
{\smallbf\ \ Table 3.}{\small\
{\it Spitzer}-IRAC photometry.\lstrut}}
\begin{tabular}{c c c c c c}        % centered columns (6 columns)
\hline\hline                 % inserts double horizontal lines
Date &                      [3.6]      &       [4.5]        &        [5.8]        &         [8.0]         &        Note \hstrut\lstrut \\
\hline
2004/03/31 & 10.67(0.03) &   9.72(0.02)  &   8.97(0.02) & 7.81(0.02) &  Trifid AOR 6049024 \hstrut \\
2005/09/21 & 10.53(0.02)  &   9.62(0.02)  &  8.89(0.02) & 7.79(0.02) &  Glimpse AOR 683519 \lstrut \\
2006/04/28 & 11.11(0.03)  &  10.02(0.03) &  9.20(0.03)  & 7.97(0.03) &   Glimpse AOR 682970 \lstrut \\ 
\hline 
\end{tabular}
}
\end{center}
\vskip 2mm
\end{table}

%%%%%%%%%%%%%%%%%%% table 4
\begin{table}[!t]
\begin{center}
\vbox{\footnotesize\tabcolsep=3pt
\parbox[c]{124mm}{\baselineskip=10pt
{\smallbf\ \ Table 4.}{\small\
{\it Spitzer}-MIPS Photometry.\lstrut}}
\begin{tabular}{c c c c }        % centered columns (6 columns)
\hline\hline                 % inserts double horizontal lines
Date & [24] & [8-24] & Note \hstrut\lstrut \\
2004/04/11 & 1.94 (0.09) & 5.87 & Trifid; Rho et al. 2006 \hstrut \\
2005/09/27 & 2.77 (0.10) & 5.02 & PID 20597 \lstrut\\
\hline
\end{tabular}
}
\end{center}
\vskip +2mm
\end{table}

%----------------------------------------------------------- fig.6
\begin{figure}
%\centering
\includegraphics[width=10.0cm]{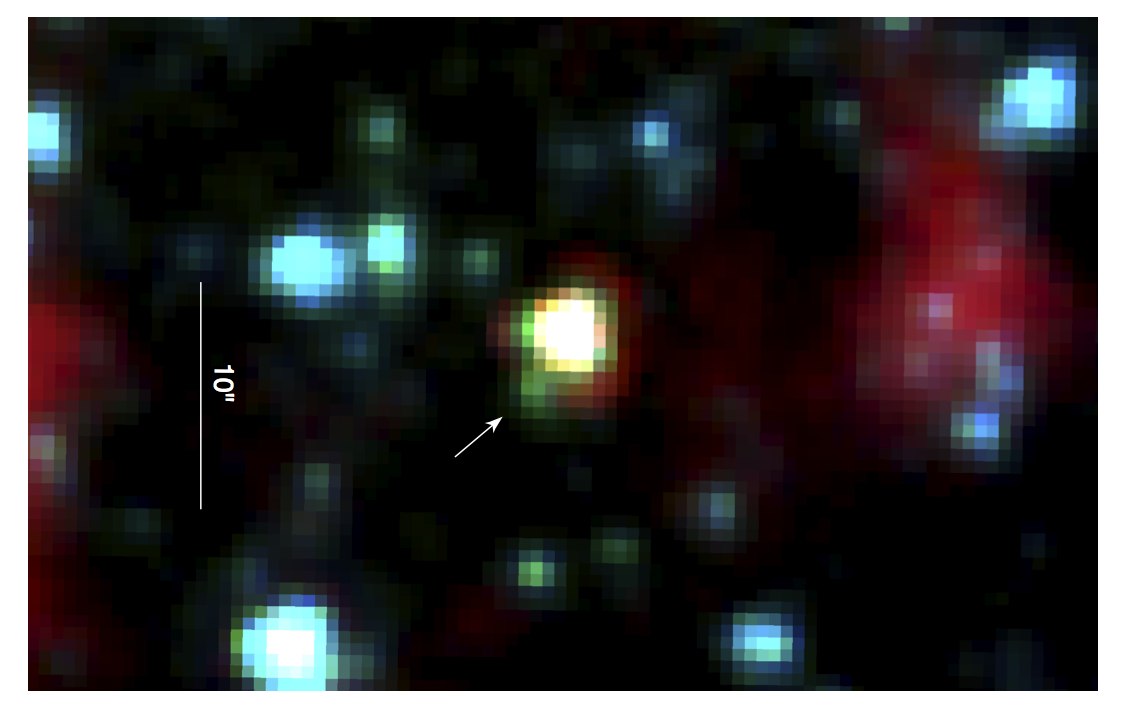}
\captionb{6}{A color composite of the 3.6 (blue), 4.5 (green) and 8.0 (red) $\mu$m images from {\it Spitzer}/IRAC of V2282~Sgr. A faint extension of the source in the SE 
direction (arrow) is evident mainly in the 4.5 $\mu$m filter, where a major contribution from H$_2$ emission is expected.}
%\label{composite}
\end{figure}
%
%______________________________________________________________

Finally we derived the photometry at 24\,$\mu$m of V2282 Sgr with an aperture of 14.3" using two different {\it Spitzer}/MIPS images: the results are shown in 
Table 4 together with the [8]$-$[24] color. The observed very red color ($>$5) in both epochs classify the star as a Class 0/I according to Rho et al. 2006). 
A variability of about 1 mag is present also at 24\,$\mu$m.

\sectionb{3}{DISCUSSION AND CONCLUSIONS}

The historic photographic $I$ passband light curve indicates that  V2282~Sgr looks like an irregular variable. An FFT analysis does not show any prominent frequency, 
as would be the case if the star were an eclipsing binary or if the variations would be due to hot or cool spots (see, e.g. Frasca et al. 2009).  
The more accurate CCD observations in $R$ passband confirm the amplitude variability of about 1 mag, with time scales of about 10 days. 

The spectrum of the star shows prominent hydrogen emission lines, and the strongest absorption lines normally found in stars of G-K spectral type. The strong absorption of the {Na}\small{I}~D lines witness the presence of a large amount of circumstellar matter, typical for young stellar objects of the class 0/I.  The high excitation [{O}\small{III}]\,$\lambda$4959 and $\lambda$5007 \AA\ ~lines also indicate the presence of surrounding matter, probably ionized by the central star of M20 (HD 164492).

To check the spectral classification based on the BFOSC spectrum and
to gain information about the stellar photosphere, we fitted the optical Spectral Energy Distribution (SED) with low-resolution synthetic spectra calculated 
with the NextGen model atmospheres (Hauschildt et al. 1999). We fixed the star distance and extinction to the values $d=1.7$\,kpc and $A_V$=\,1.3\,mag reported by 
Lynds et al. (1985) and we let $T_{\rm eff}$ and $\log g$ free to vary, adopting a standard extinction law $A_V=3.1\times E(B-V)$. The fit was limited to the photometric 
bands from $B$ to $I$, because afterwards the photospheric flux could be significantly contaminated by an IR excess.
The best-fit model gives $\log g=3.5$ and $T_{\rm eff}=4400$\,K. With these parameters, a radius of about 4.5\,$R_{\sun}$ would be derived for the star. However, the star temperature is strongly dependent on the assumed extinction. The star is probably located in the near side of the M20 nebula, given its low reddening. 

The {\it Spitzer} IRAC and MIPS observations reported in the previous Section show that V2282 Sgr has an accretion disk. This is apparently in contrast with the lack of a near-IR excess observed at several epochs, but present in the UKIDSS images. Although we do not have a dense coverage of near-IR observations, we think that the variability could be due to a combination of variable extinction and changes in the inner accretion disk (Rice et al. 2012).
Combining the  {\it Spitzer} IRAC and MIPS observations with the UKIDSS JHK$_s$ data taken in 2006,  when the star showed a near-IR excess, we obtained the infrared SED of V2282 Sgr. This SED has been fitted with the infalling envelope+disk+central source radiation transfer model described by Robitaille et al. (2006), using the fitting tool by Robitaille et al. (2007), see  Figure~7. The best fit model parameters are collected in Table 5. Of course, these parameters are not very exact because the near- and mid-IR observations were not simultaneous. 
Taking into account the source flux and color variability, the agreement between the stellar source parameters  derived from the optical  and IR SEDs is acceptable. 
The presence of a non-homogeneous circumstellar disk is also suggested by optical variability on short time scale, as indicated by the dense CCD sampling in 2009, which shows some episodes of fast dimming with a time scale of about 10 days.

Also the X-ray luminosity (7.6 10$^{32}$ erg/s) derived from the {\it Chandra} count rate, assuming a thermal spectrum and the adopted distance, is compatible with a PMS star of 
1$\le M \le 2M_{\sun}$ (Preibisch 2007), possibly of Class I, associated with an accreting circumstellar disk.

%----- Table 5
\begin{table}[!t]
\begin{center}
\vbox{\footnotesize\tabcolsep=3pt
\parbox[c]{124mm}{\baselineskip=10pt
{\smallbf\ \ Table 5.}{\small\
Physical parameters of V2282 Sgr.\lstrut}}}
%\centering                          % used for centering table
\begin{tabular}{l l }        % centered columns (2 columns)
\hline\hline                 % inserts double horizontal lines
Stellar mass & 1.7 $M_{\sun}$\\
Stellar temperature & 4270 K\\
Envelope accretion rate & 1.6$\times 10^{-5}$ $M_{\sun} / yr$\\
Envelope outer radius & 5100 AU \\
Envelope cavity angle & 22$\fdg$5 \\
Disk mass & 0.035 $M_{\sun}$\\
Disk outer radius & 190 AU\\
Disk accretion rate & 9.7$\times 10^{-7}$ $M_{\sun} / yr$ \\
$A_V$ & 2.58 mag \\
Distance & 1.9 kpc \\
\hline
\end{tabular}
\end{center}
\end{table}
%---
%-------------------------------- fig.7
\begin{figure}
%   \centering
 \includegraphics[width=9cm]{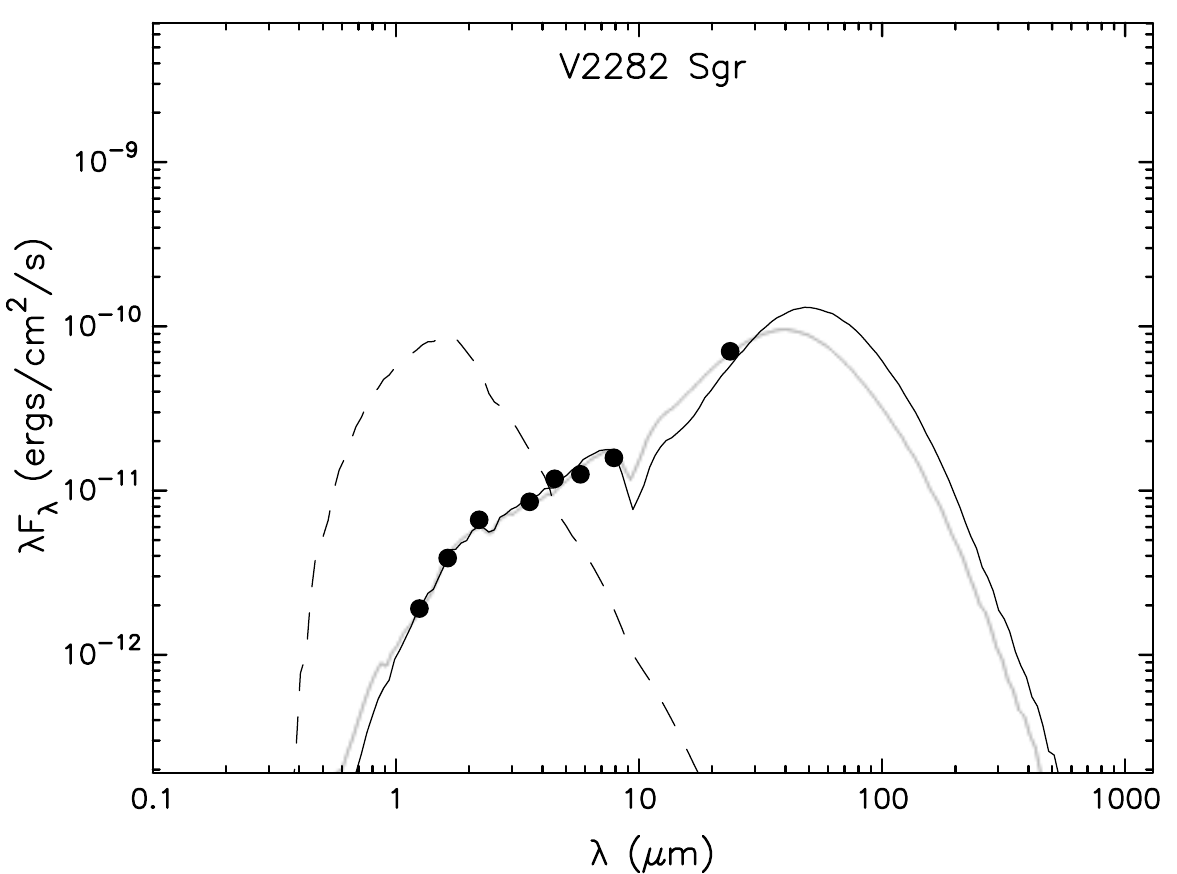}
      \captionb{7}{SED of V2282~Sgr from MIPS, IRAC and UKIDSS data (dots). The best-fitting disk model (Robitaille 2007) is over-plotted to the data with a continuous line. The dashed line is the model stellar component (Table 5).
      }
%         \label{sedir}
   \end{figure}
%______________________________________________________________

We suggest that the infrared emitting dust torus surrounding the star,  is seen pole-on, so that we can see the inner source without strong absorption.  Similar cases have already been pointed out in the literature (DK Cha, see Garcia et al. 2011).
%%%%%%%%%%%%%%%%%%%

\smallskip

\thanks{
This research was supported by funds of the University La Sapienza and of INAF-OACT. 
R.N. thanks the Direction of the Asiago Observatory for hospitality during the plate scans. 
The authors thank the Bologna Observatory for the logistic support and the technical assistance during the observations. We acknowledge the use of astronomical catalogs  from CDS (Strasbourg), DENIS and NASA/IPAC data archive. }

%%%%%%%%%%%%%%%%%%%%%%%

\smallskip

\References

\refb  Cambresy L., Rho, J., Marshall, D.,J., Reach, W.T. 2011, A\&A, 527, 141.
\refb  Cohen, M. \& Kuhi, M.V. 1979, ApJS, 41, 743.
\refb  Cornero N., http://obswww.unige.ch/~behrend/pagedcou.html
\refb  Cutri, R. M. Skrutskie, M. F., van Dyk, S. et al. 2003, The IRSA 2MASS All-Sky Point Source Catalog, NASA/IPAC Infrared Science Archive.
\refb  Cutri, R. M. Skrutskie, M. F., van Dyk, S. et al. 2012, The WISE All-Sky Point Source Catalog. CDS catalog II/311.
\refb  Duncan, J.C. 1921, PASP, 33, 207
\refb  Frasca, A., Covino, E., Spezzi, L., et al. 2009, A\&A, 508, 1313
\refb  Garcia Lopez, R., Nisini, B., Antoniucci, S., et al. 2011, A\&A, 534, 99
\refb  Giannini, T.,  Lorenzetti, D., D'Elia D. 2009, ApJ, 704, 606
\refb  Glasby, J.S., The nebular variables, Pergamon Press, 1974
\refb Hauschildt, P. H., Allard, F., Ferguson, J., Baron, E., et al. 1999, ApJ, 525, 871
\refb Herbig, G.H. 1957, ApJ, 125, 654
\refb  Kaas, A.A. 1999, AJ, 118, 558
\refb  Kharchenko, N.V. Piskunov, A.E., Roser, S., Schilbach, et al. 2005, A\&A, 438, 1163
\refb  Kohoutek, L., Mayer, P., Lorenz, R. 1999, A\&AS, 134, 129
\refb  Kukarkin, B.V., et al. 1968, IBVS 311
\refb  Lampland, C.O. 1919, Pop. Astr. 27, 32
\refb  Lampland, C.O. 1923, Pub. A.A.S. 4, 30
\refb  Lynds, B. T., Canzian, B. J., O'Neil, E. J., Jr. 1985, ApJ, 288, 164
\refb  Maffei, P., 1959, MemSAIt 30, 83
\refb  Maffei, P., 1963, MemSAIt 34, 441
\refb  Morales-Calder\'on, M., Stauffer, J. R., Rebull, L., et al. 2009, ApJ, 702, 1507
\refb Nesci, R., Massaro, E. Rossi, C., Sclavi, S., AJ, 2006, 130, 1466
\refb  Noriega-Crespo, A., Moro-Martin, A., Carey, S. et al., 2004, ApJS 154, 402
\refb Ogura, K., \& Ishida, K. 1975, PASJ, 27, 119
\refb Preibisch, T. 2007, MemSAIt 78, 1 
\refb  Rho, J., Reach, W.T, Lefloch, B., Fazio, G.G, 2006, ApJ,  643, 965
\refb  Rho, J., Ramirez, S.V., Corcoran, M.F., Hamaguchi, K., et al. 2004, ApJ, 607, 904 
\refb  Rice, T. S., Wolk, S. J., Aspin, C. 2012, ApJ, 755, 65
\refb  Robitaille, T. P., Whitney, B.A., Indebetouw, R., et al., 2006, ApJS 167, 256
\refb  Robitaille, T. P., Whitney, B.A., Indebetouw, R., Wood, K., 2007, ApJ 169, 328 
\refb  Robitaille, T. P., Meade, M. R., Babler, B. L. et al. 2008, AJ, 136, 2413
\refb  Samus, N.N., Durlevich, O.V., Kazarovets E V., et al., 2012, General Catalog of Variable Stars, CDS catalog B/gcvs
\refb Wright, E.L., Eisenhardt, P.R.M., Mainzer, A.K., et al., 2010 AJ, 140, 1868 
\refb Ybarra, J. E. \& Lada, E. A. 2009, ApJ, 695, 120
\refb  Yusef-Zadeh, F., Biretta, J, Geballe, T.R., 2005, AJ, 130, 1171

%%%%%%%%%%%%%%%%%%%

\end{document}